\documentclass[12pt,a4paper]{article}

\pdfoutput=1

\usepackage[a4paper]{geometry}
\usepackage{graphicx}
\usepackage{amsmath}
\usepackage{amssymb}
\usepackage{float}
\usepackage{listings}
\title{\bf Inclusive three jet production at the LHC as a new BFKL probe}
\author{F. Caporale$^1$,  G. Chachamis$^1$, 
       B.Murdaca$^2$,  A. Sabio Vera$^{1}$\\ \\
{\small $^1$ Instituto de F{\' \i}sica Te{\' o}rica UAM/CSIC, Nicol{\'a}s Cabrera 15}\\ 
{\small \& Universidad Aut{\' o}noma de Madrid, E-28049 Madrid, Spain.}\\
{\small $^2$ Dipartimento di Fisica, Universit{\`a} della Calabria \&}\\
{\small Istituto Nazionale di Fisica Nucleare, Gruppo Collegato di Cosenza,}\\
{\small I-87036 Arcavacata di Rende, Cosenza, Italy.}
}

\begin{document}

\maketitle 

\section{Introduction}

The Balitsky-Fadin-Kuraev-Lipatov (BFKL) formalism~\cite{Balitsky:1978ic,Kuraev:1977fs,Kuraev:1976ge,Fadin:1998py,Ciafaloni:1998gs}
is one of the most important resummation programmes
in high energy QCD.  
A typical BFKL observable at the LHC is the azimuthal angle ($\phi$) decorrelation of two tagged forward/backward jets widely separated in rapidity, $Y$, in the so-called Mueller-Navelet jets setup~\cite{Mueller:1986ey}. This multiple emission  appears as a fast decrease of $\langle \cos{(n \, \phi)} \rangle$ as a function of $Y$~\cite{DelDuca:1993mn,Stirling:1994he,Orr:1997im,Kwiecinski:2001nh}. However, these differential distributions suffer from a large influence of collinear regions in phase space, so it was proposed to remove the $n=0$ dependence by studying the ratios ${\cal C}_{m,n} = \langle \cos{(m \, \phi)} \rangle / \langle \cos{(n \, \phi)} \rangle$~\cite{Vera:2006un,Vera:2007kn,Angioni:2011wj}.  In recent studies~\cite{Ducloue:2013bva,Caporale:2014gpa}, a BFKL analysis at NLL is able to fit the large $Y$ tail of the Mueller-Navelet  ${\cal C}_{m,n}$  ratios.

In Ref.~\cite{Caporale:2015vya}, we proposed new observables related to final states with two tagged forward jets separated by a large rapidity span, along with a third tagged jet produced in the central region of rapidity, allowing for inclusive radiation in the remaining areas of the detectors.

The two tagged forward jets $A$ and $B$ have transverse momentum $\vec{k}_{A,B}$, azimuthal angle 
$\theta_{A,B}$ and rapidity $Y_{A,B}$. The central jet is characterized by $\vec{k}_J$, 
$\theta_J$ and $y_J$ and the differential cross section on these variables can be written in the form
\begin{eqnarray}
\frac{d^3 \sigma^{3-{\rm jet}}}{d^2 \vec{k}_J d y_J}   &=& 
 \frac{\bar{\alpha}_s }{\pi k_J^2}   \int d^2 \vec{p}_A \int d^2 \vec{p}_B \, 
 \delta^{(2)} \left(\vec{p}_A + \vec{k}_J- \vec{p}_B\right)
 \\ \nonumber
&&
 \times \,\, \varphi \left(\vec{k}_A,\vec{p}_A,Y_A - y_J\right) 
 \varphi \left(\vec{p}_B,\vec{k}_B,y_J - Y_B\right)
 \label{Onejetemission}
\end{eqnarray}
where we assume that $Y_A > y_J > Y_B$ and $k_J$ lies above the experimental resolution 
scale. $\varphi$ are BFKL gluon Green functions normalized to $ \varphi \left(\vec{p},\vec{q},0\right) = \delta^{(2)} \left(\vec{p} - \vec{q}\right)$ and $\bar{\alpha}_s = \alpha_s N_c/\pi$. Now, new distributions can be defined using the projections on the two relative azimuthal angles formed by each of the forward jets with the central jet, $\theta_A - \theta_J - \pi$ and $\theta_J - \theta_B - \pi$:
\begin{eqnarray}
&&\hspace{-1cm}\int_0^{2 \pi} d \theta_A \int_0^{2 \pi} d \theta_B \int_0^{2 \pi} d \theta_J \cos{\left(M \left( \theta_A - \theta_J - \pi\right)\right)} 
\cos{\left(N \left( \theta_J - \theta_B - \pi\right)\right)}
\frac{d^3 \sigma^{3-{\rm jet}}}{d^2 \vec{k}_J d y_J} 
\nonumber\\
&&\hspace{-1cm}= \bar{\alpha}_s \sum_{L=0}^{N} 
\left( \begin{array}{c}
\hspace{-.2cm}N \\
\hspace{-.2cm}L\end{array} \hspace{-.18cm}\right)
 \left(k_J^2\right)^{(L-1/2)} 
\int_{0}^\infty d p^2 \, \left(p^2\right)^{(N-L/2)} 
\int_0^{2 \pi}  d \theta    \frac
{   (-1)^{M+N} \cos{ \left(M \theta\right)} \cos{\left((N-L) \theta\right)}
}{
 \sqrt{\left(p^2 + k_J^2+ 2 \sqrt{p^2 k_J^2} \cos{\theta}\right)^{N}}
}\nonumber\\
&&
\times \, \, \phi_{M} \left(p_A^2,p^2,Y_A-y_J\right)
\phi_{N} \left(p^2+ k_J^2 + 2 \sqrt{p^2 k_J^2}\cos{\theta},p_B^2,y_J-Y_B\right),
\end{eqnarray}
with $\phi_{n}$ defined in Ref.~\cite{Caporale:2015vya}.
The experimentally relevant observable is the mean value in the selected events of the two cosines, {\it i.e.}
\begin{eqnarray}
\langle \cos{\left(M \left( \theta_A - \theta_J - \pi\right)\right)}  
\cos{\left(N \left( \theta_J - \theta_B - \pi\right)\right)}
\rangle && \\
&&\hspace{-8.5cm} = \frac{\int_0^{2 \pi} d \theta_A d \theta_B d \theta_J \cos{\left(M \left( \theta_A - \theta_J - \pi\right)\right)}  \cos{\left(N \left( \theta_J - \theta_B - \pi\right)\right)}
\frac{d^3 \sigma^{3-{\rm jet}}}{d^2 \vec{k}_J d y_J} }{\int_0^{2 \pi} d \theta_A d \theta_B d \theta_J 
\frac{d^3 \sigma^{3-{\rm jet}}}{d^2 \vec{k}_J d y_J} }.\nonumber
\end{eqnarray}
In order to have optimal perturbative convergence and eliminate collinear contamination, we can remove the contributions from zero conformal spin by defining the ratios:
\begin{eqnarray}
{\cal R}_{P,Q}^{M,N} &=& \frac{\langle \cos{\left(M \left( \theta_A - \theta_J - \pi\right)\right)}  
\cos{\left(N \left( \theta_J - \theta_B - \pi\right)\right)}
\rangle}{\langle \cos{\left(P \left( \theta_A - \theta_J - \pi\right)\right)}  
\cos{\left(Q \left( \theta_J - \theta_B - \pi\right)\right)}
\rangle}, \,\,\, 
\label{RMNPQ}
\end{eqnarray}
with $M, \,N, \, P, \, Q \, = \, 1,\, 2$.

\begin{figure}
\begin{center}
   \includegraphics[height=6.cm]{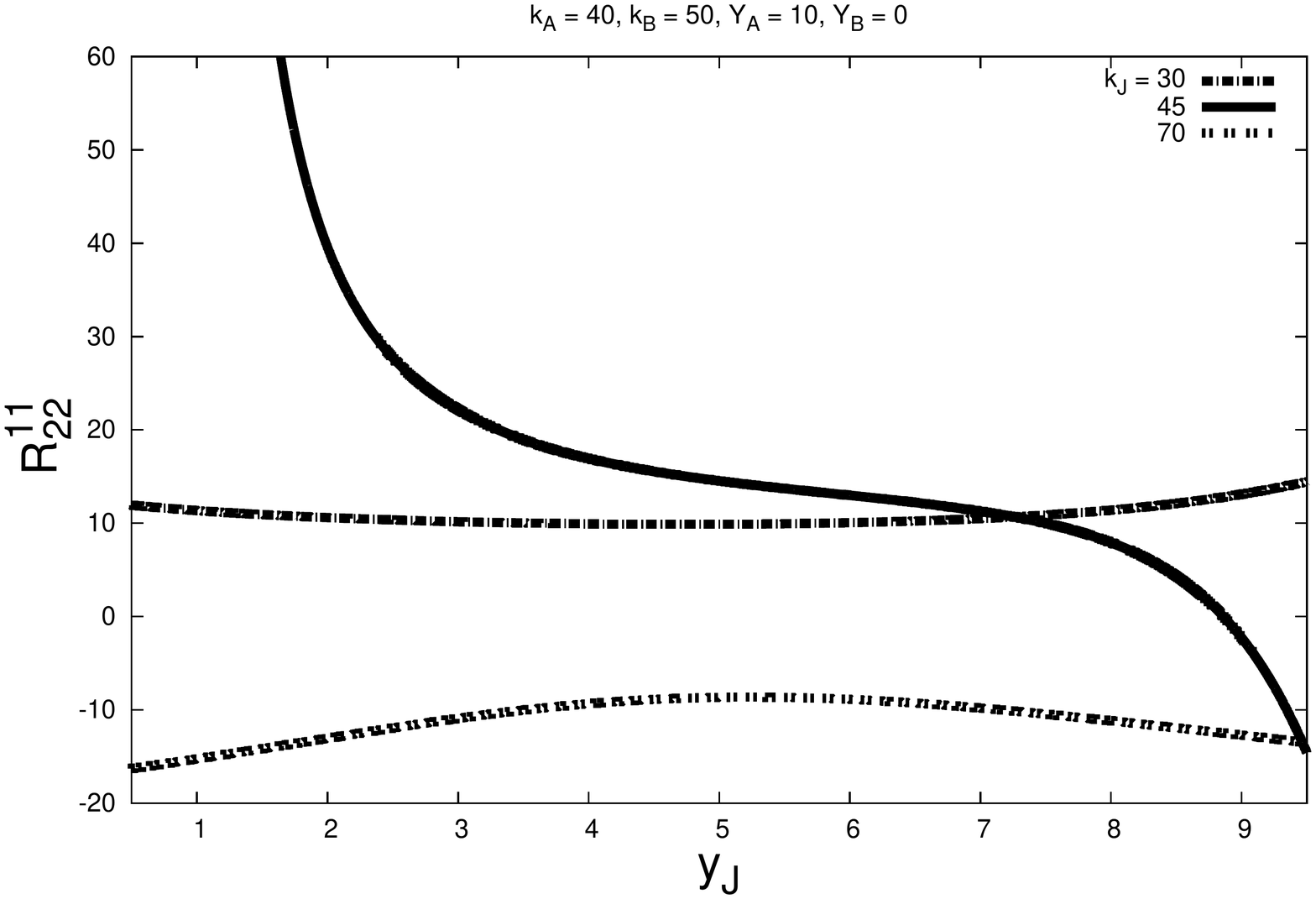}
   \includegraphics[height=6.cm]{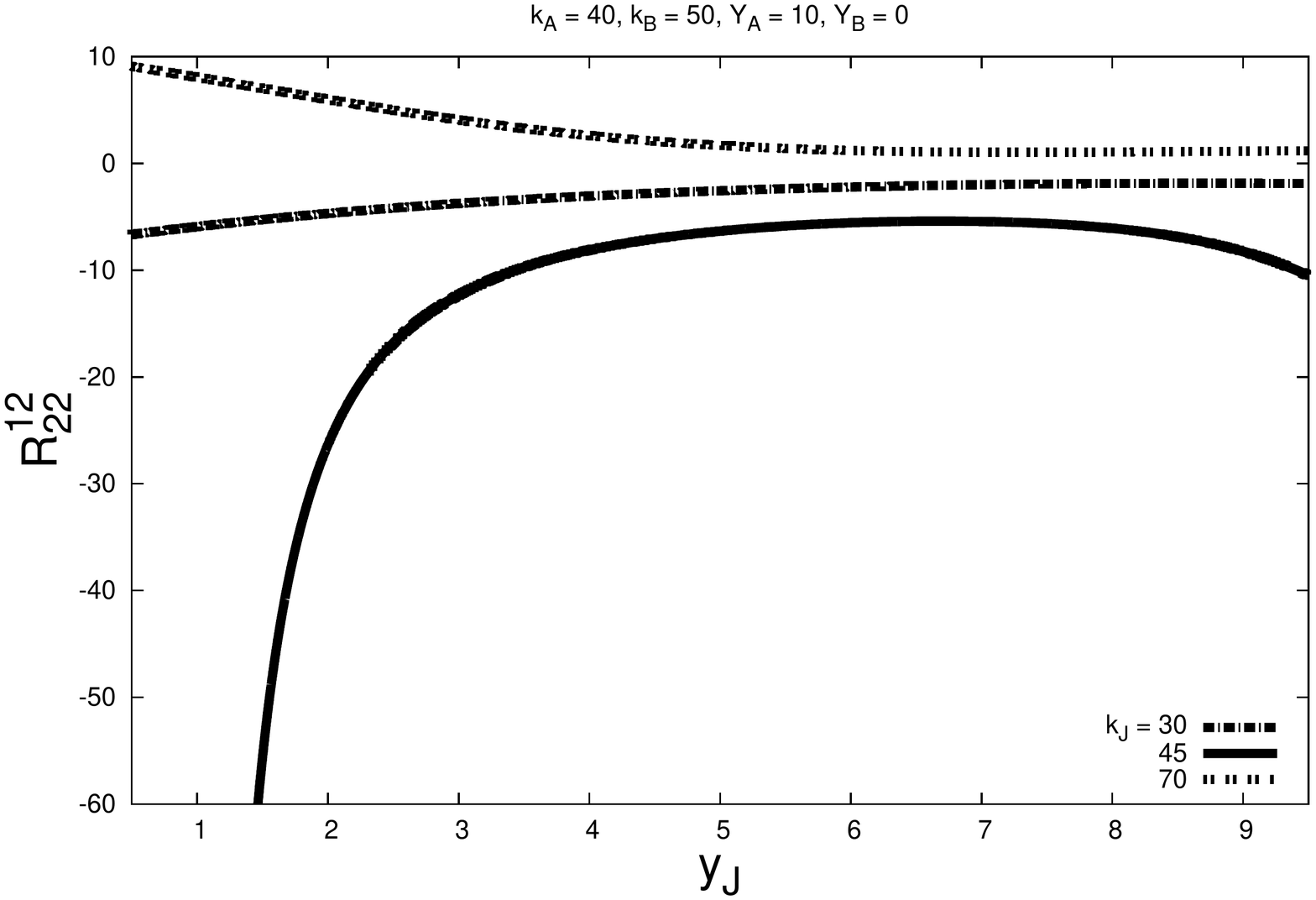}
\end{center}
\caption{The ratios ${\cal R}_{2,2}^{1,1}$ and ${\cal R}_{2,2}^{1,2}$  as a function of the rapidity of the central jet $y_J$.}
\label{RMNPQfigure}
\end{figure}

\section{Results and Outlook}
Now, different momenta configurations can be investigated. Here, in Fig.~\ref{RMNPQfigure},
two ratios ${\cal R}_{P,Q}^{M,N}$ with $M,N=1,2$ are shown while the momenta of the forward jets are fixed to $k_A=40$ GeV and $k_B=50$ GeV and 
their rapidities also fixed to $Y_A=10$ and $Y_B=0$. The central jet transverse momentum
takes three values, $k_J= 30, 45, 70$ GeV and  the rapidity of the central jet $y_J$ 
is varied in between the forward/backward jet rapidities. 

It will be very interesting to see the predictions from fixed order analyses as well as from the BFKL inspired Monte Carlo {\tt BFKLex}~\cite{Chachamis:2011rw,Chachamis:2011nz,Chachamis:2012fk,
Chachamis:2012qw,Caporale:2013bva,Chachamis:2015zzp,Chachamis:2015ico} for these and other 
similar observables~\cite{Caporale:2015int,Caporale:2016} where the projection on azimuthal angles 
is used. This type of observables  will be crucial to define the region of phenomenological applicability of the BFKL resummation.

\section*{Acknowledgements}
G.C. acknowledges support from the MICINN, Spain, under contract FPA2013-44773-P. 
A.S.V. acknowledges support from Spanish Government (MICINN (FPA2010-17747, FPA2012-32828)) and, together with F.C and B.M., to the Spanish MINECO Centro de Excelencia Severo Ochoa Programme (SEV-2012-0249). The work of B.M. was supported in part by the grant RFBR-13-02-90907 and by the European Commission, European Social Fund and Calabria Region, that disclaim any liability for the use that can be done of the information provided here.

\bibliographystyle{MPI2015}
\bibliography{chachamis-refs}

\end{document}